# Mutual Information in Coupled Double Quantum Dots: A Simple Analytic Model for Potential Artificial Consciousness

Katsuaki Tanabe*

Department of Chemical Engineering, Kyoto University, Nishikyo, Kyoto 615-8510, Japan
* Correspondence: tanabe@cheme.kyoto-u.ac.jp

**Abstract:** The integrated information theory is thought to be a key clue towards the theoretical understanding of consciousness. In this study, we propose a simple numerical model comprising a set of coupled double quantum dots, where the disconnection of the elements is represented by the removal of Coulomb interaction between the quantum dots, for the quantitative investigation of integrated information. As a measure of integrated information, we calculate the mutual information in the model system, as the Kullback–Leibler divergence between the connected and disconnected status, through the probability distribution of the electronic states from the master transition-rate equations. We reasonably demonstrate that the increase in the strength of interaction between the quantum dots leads to higher mutual information, owing to the larger divergence in the probability distributions of the electronic states. Our model setup could be a useful basic tool for numerical analyses in the field of integrated information theory.

## 1. Introduction

Unveiling of the origin and quantification of consciousness in the brains of living beings and potentially even in the artificial functional machines in relation to statistical mechanics is of significant interest. The integrated information theory [1–4] is thought to be a key clue towards the theoretical understanding of consciousness [5–10]. In the field of integrated information theory, several quantitative measures to representatively evaluate the degree of consciousness have been proposed so far: mutual information [1], transfer entropy [11], stochastic interaction [12,13], and geometric integrated information [14]. Among these measures of integrated information, the mutual information is positioned as the simplest but still applicable to quantify the information loss caused by splitting a system into parts, particularly for the cases of bi-partition [15]. By contrast, it is generally difficult to analytically solve the minimization of the Kullback–Leibler divergence to determine the values of the other measures of integrated information in practical systems due to their non-submodularity [14–16]. Furthermore, among the proposed measures of integrated information, the mutual information can solely be applied for steady-state systems, while all the other measures are defined in the time-evolving framework through the past to the future.

It is postulated in the integrated information theory that the consciousness arises from the situation where multiple elementary information are tightly connected with one another, and one simultaneously recognizes the entire connected system as a whole. Hence, it may be seemingly natural to consider that the intensity of consciousness could be quantified as the difference in the amount of information between the connected and disconnected systems, which can be represented by the Kullback–Leibler divergence between their probability distributions. This context is our recognition through References 1–16 for the relationship between the effective information discussed in the initial concept of the integrated information theory by Tononi [1] and the Kullback–Leibler divergence or the measures of integrated information listed above. While the mathematical setup to formulate the integrated information has been established to some extent [1,2,8,9,11–14] and implementations of the numerical to practical complex data sets have been demonstrated [2,5–10,15,16], our present work is positioned as a provision of a simple numerical model setup that can be used in a wide range of applications spanning from an introductory educational material to a basic prototype tool of detailed numerical investigations for further scholarly understanding in the field of integrated information theory.

Bipartite, four-state configurations are a handy model employed for the investigations in the field of information thermodynamics [17–22]. Quantum dots [23–27], often referred to as artificial atoms, are an adopted candidate for a material component of such setups [17,28–33]. The discretized density of states in quantum dots, owing to the three-dimensional confinement of electrical carriers, enables single-electron manipulation [34–36], which may make the discussions clear in information statistical mechanics. We previously studied a coupled double quantum dot system as an autonomous information engine, calculating the steady-state entropy production rate in each component, heat and electron transfer rates via the probability distribution of the four electronic states from the master transition-rate equations, to acquire device-design principles toward the realization of corresponding practical energy converters [37]. In the present study, we employ the coupled double quantum dot system to demonstrate case-study simulations of mutual information between the quantum dots, to provide a simple but practically useful model setup for numerical investigations of the degree of consciousness based on the integrated information theory.

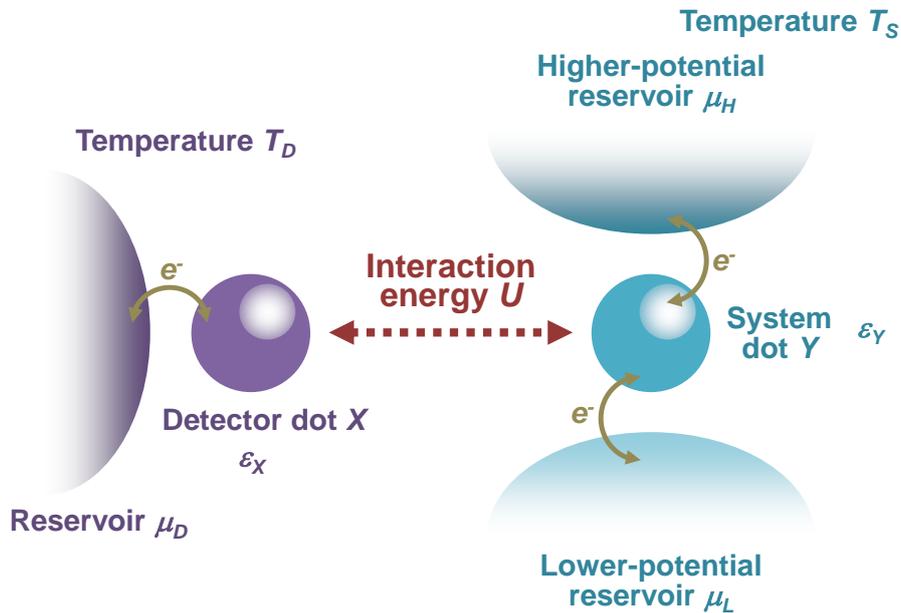

**Figure 1.** Schematic illustration of the model coupled double quantum dot system.

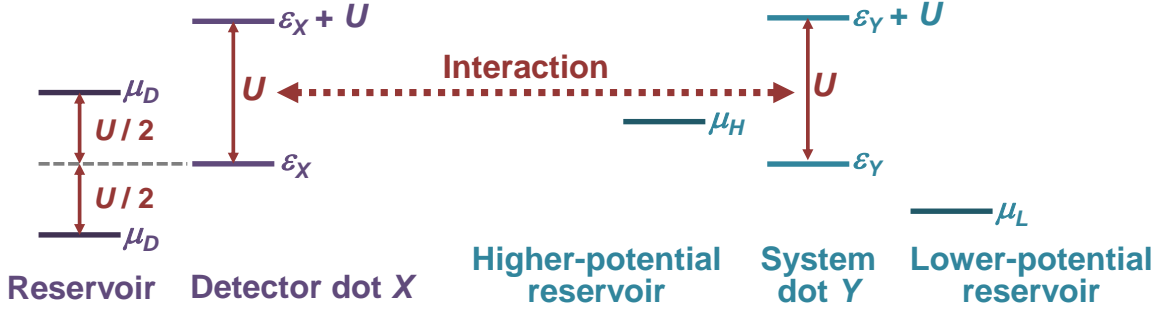

**Figure 2.** Schematic diagram of the relationship among the potential energy levels in the components of the coupled double quantum dot system model setup.

## 2. Theory and Calculation Methods

The model setup of the present study comprises two quantum dots and three thermal/electronic reservoirs around as schematically depicted in Figure 1. Each quantum dot can contain up to one electron. One quantum dot with an electronic potential energy $\varepsilon_X$ functions as an electronic detector by "checking" whether an electron is in the other quantum dot through capacitive interaction strength or Coulomb interaction energy $U$ between the two quantum dots. This "detector dot" is kept at a temperature $T_D$ and connected to thermal and electronic reservoirs, both having the same temperature $T_D$ and an electronic potential energy $\mu_D$. The other quantum dot with an electronic potential energy $\varepsilon_Y$ is connected to two reservoirs through electrical leads and enables an electrical current flow. This "system dot" is kept at a temperature $T_S$ and connected to two thermal and electronic reservoirs both at $T_S$ with electronic potential energies $\mu_H$ and $\mu_L$ ($\mu_H > \mu_L$). It should be noted that by such a connecting configuration, this quantum-dot system realizes a non-equilibrium steady state, and is productively active, performing work in exchange for the decrease in thermal energy (the *thermoelectric engine* mode operation) or the increase in entropy (the *environmental information engine* mode operation) of the reservoirs, as living bodies [37]. The potential-energy relations among the components in the setup are schematically shown in Figure 2 for clarification. This double-quantum-dot configuration as a whole can drive electrical current between the reservoirs of $\mu_H$ and $\mu_L$ through the system dot, even in the direction from $\mu_L$ to $\mu_H$ against the potential slope and thus generate work by properly setting the transition or tunneling rates across the interfaces between the quantum dots and reservoirs, as shown in Reference 37. Each quantum dot has an electronic state 0 or 1, where 1 and 0 mean that the dot is filled or not filled (i.e., empty) with an electron, respectively. In this way, the total electronic state $(x, y)$ will be $(0, 0)$, $(0, 1)$, $(1, 0)$, or $(1, 1)$. For the state $(1, 1)$, the electronic potential energy in the quantum dots will increase to $\varepsilon_X + U$ and $\varepsilon_Y + U$ for the detector and system dots, respectively, due to Coulomb repulsion. We set the time resolution fine enough so that no simultaneous or diagonal jump, such as a transition from $(0, 0)$ to $(1, 1)$, is assumed in our bipartite setup.

The time evolution of the probability of the state $p(x, y)$ can be generally written as a master equation:

$$d_t p(x, y) = \sum_{x', y'} \left\{ W_{x,x'}^{y,y'} p(x', y') - W_{x',x}^{y',y} p(x, y) \right\}, \quad (1)$$

where $W_{x,x'}^{y,y'}$ is the transition rate from a state $(x', y')$ to $(x, y)$ and we have:

$$W_{10}^{y} = \Gamma f_y, \quad (2)$$

and

$$W_{01}^y = \Gamma(1 - f_y) \quad (3)$$

for the electron transfer on the detector dot and

$$W_x^{10,\upsilon} = \Gamma_x^\upsilon f_x^\upsilon, \quad (4)$$

and

$$W_x^{01,\upsilon} = \Gamma_x^\upsilon (1 - f_x^\upsilon) \quad (5)$$

for the system dot in this model. Note again that for the jumps, either $x$ or $y$ is fixed at each time step. $\Gamma$ is the electronic tunneling rate between the detector dot and its reservoir. We assume the density of states in the detector-side reservoir to be uniform so that $\Gamma$ is independent of $y$. $\Gamma_x^\upsilon$ is the tunneling rate between the system dot and its reservoirs where $\upsilon = H$ or $L$ corresponds to the higher- or lower-potential reservoir, respectively. In contrast, we assume nonuniform profiles of the density of states in the system-side reservoirs so that $\Gamma_x^\upsilon$ depends on $x$. Fermi distribution functions for the detector and system dots have forms of:

$$f_y = \frac{1}{1 + \exp\left(\frac{\varepsilon_X + yU - \mu_D}{T_D}\right)} \quad (6)$$

and

$$f_x^\upsilon = \frac{1}{1 + \exp\left(\frac{\varepsilon_Y + xU - \mu_\upsilon}{T_S}\right)}, \quad (7)$$

respectively. For simplicity, the Boltzmann constant is set to unity or absorbed into the temperatures throughout this paper. We then determine the steady-state probability distribution of $p(x, y)$ for the four electronic states through the master transition-rate equations of Equation (1).

For the disconnected model, we consider the probability distribution of the states $q(x, y)$ in the system under consideration that corresponds to $p(x, y)$ for the case $U = 0$, where the interactive information between the quantum dots is removed. We then define the mutual information in the coupled double quantum dot system, via the Kullback–Leibler divergence, as:

$$\Phi_{MI} = D_{KL}\{p(x,y) \| q(x,y)\} = \sum_{x,y} p(x,y) \ln \frac{p(x,y)}{q(x,y)}$$
$$= p(0,0) \ln \frac{p(0,0)}{q(0,0)} + p(0,1) \ln \frac{p(0,1)}{q(0,1)} + p(1,0) \ln \frac{p(1,0)}{q(1,0)} + p(1,1) \ln \frac{p(1,1)}{q(1,1)}, \quad (8)$$

where $D_{KL}\{p(x,y) \| q(x,y)\}$ is the Kullback–Leibler divergence between the probability distributions $p(x, y)$ and $q(x, y)$.

**3. Results and Discussion**

Figure 3 presents a set of our calculation results for the mutual information of the coupled double quantum dot system in dependence on the capacitive interaction strength or quantum-dot Coulomb repulsion energy $U$ for varied temperatures of the detector side $T_D$, for the cases $\mu_D = \varepsilon_X - U/2$ and $\mu_D = \varepsilon_X + U/2$ (adopted from Reference 37). It should be noted that, because there are too many conditional parameters to change for the numerical calculations, we intentionally provided some restrictions among the parameters in this study. It is observed for all $T_D$'s that as $U$ increases, the mutual information increases. This resulted trend seems reasonable in the view of the integration of information in the system; stronger interaction between the quantum dots provides a higher degree of integrated information. The mutual information is observed to be strongly dependent on $T_D$ or the relative difference between $T_S$ and $T_D$. Lower $T_D$ provides higher mutual information for a given $U$. This trend can be attributed to the fact that under the fixed-$T_S$ conditions, lower $T_D$ and thus larger difference between $T_S$ and $T_D$ provides higher electron-transport selectivity, functioning towards the same directionality as larger $U$. With no interaction between the quantum dots, $U = 0$, the four probabilities of the states almost equivalently share the pie, $q(x, y) = 0.25$. In the large-$U$ regime, the repulsive interaction between the quantum dots tends to eliminate $p(1, 1)$. For the condition $\mu_D = \varepsilon_X - U/2$, for larger $U$, $p(0, 0)$ and $p(0, 1)$ approach 0.5 while the other probabilities of the states go to zero, owing to the deep reservoir level for the detector dot and $\varepsilon_Y$ locating in the middle of $\mu_H$ and $\mu_L$ for the system dot. This situation provides the limit of mutual information as $2 \cdot 0.5 \cdot \ln(0.5/0.25) = \ln 2 \sim 0.693$, as observed in Figure 3 (a). On the other hand, for the condition $\mu_D = \varepsilon_X + U/2$, for larger $U$, $p(1, 0)$ approaches unity while the others go to zero, because of the large reservoir-to-dot potential drop for the detector dot. This situation provides the limit of mutual information as $1 \cdot \ln(1/0.25) = \ln 4 \sim 1.386$, as observed in Figure 3 (b). We adopt the condition $\mu_D = \varepsilon_X + U/2$ in the following calculations for higher mutual information.

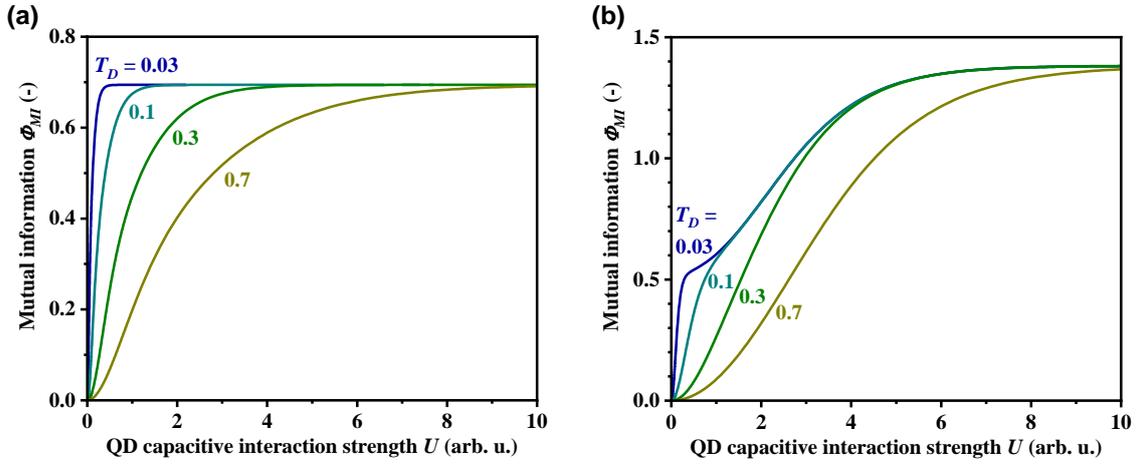

**Figure 3.** Mutual information $\Phi_{MI}$ in the coupled double quantum dot system in dependence on the capacitive interaction strength between the quantum dots $U$ for various temperatures of the detector quantum dot $T_D$ for the cases (a) $\mu_D = \varepsilon_X - U/2$ and (b) $\mu_D = \varepsilon_X + U/2$ under the condition $\varepsilon_X = \varepsilon_Y = 1$, $\mu_H = 1.1$, $\mu_L = 0.9$, $T_S = 1$, $\Gamma = 100$, $\Gamma_0^H = \Gamma_1^L = 10$, and $\Gamma_1^H = \Gamma_0^L = 0.1$. QD stands for quantum dot.

We plot the mutual information for various tunneling rates in Figure 4. We varied the tunneling rates for the electron-flow direction against the potential differences $\Gamma_0^H$ and $\Gamma_1^L$ while fixing the tunneling rates in the direction down the potential slopes $\Gamma_1^H$ and $\Gamma_0^L$. As seen in this set of mutual-information results, the mutual information is moderately dependent on the antisymmetric relative amplitudes of the electronic tunneling rates, i.e., that $\Gamma_0^H$ and $\Gamma_1^L$ are larger than $\Gamma_1^H$ and $\Gamma_0^L$. This type of antisymmetry was essential for the operation of the coupled double quantum dot system as an information engine and strongly influenced the information-engine efficiency [37]. However, it is observed that this is not the case for the mutual information, and it works even for the nonsymmetric case, $\Gamma_0^H = \Gamma_0^L = \Gamma_1^H = \Gamma_1^L = 0.1$.

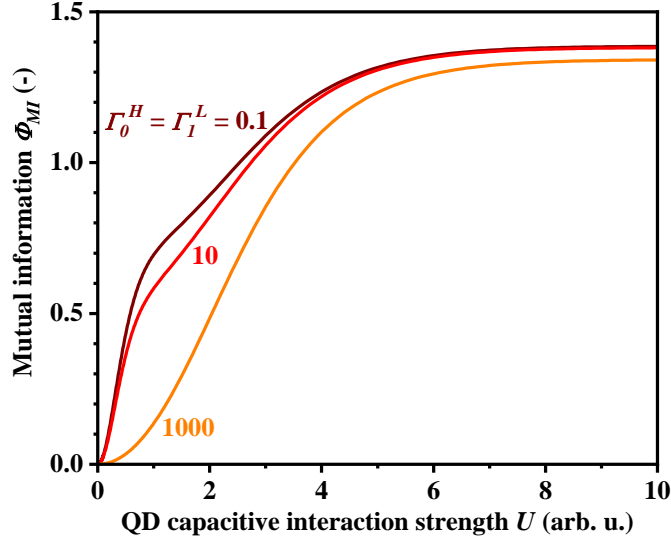

**Figure 4.** Mutual information $\Phi_{MI}$ in dependence on the capacitive interaction strength between the quantum dots $U$ for various tunneling rates $\Gamma_0^H$ and $\Gamma_1^L$, under the condition $\varepsilon_X = \varepsilon_Y = 1$, $\mu_H = 1.1$, $\mu_L = 0.9$, $T_D = 0.1$, $T_S = 1$, $\Gamma = 100$, and $\Gamma_1^H = \Gamma_0^L = 0.1$.

Here, we investigate the influence of the energy levels' relative positions among the system dot and its reservoirs. Figure 5 shows the mutual information of the coupled double quantum dot system with varied $\mu_H$ and $\mu_L$ to $\varepsilon_Y$, keeping $\mu_H - \mu_L$ constant. The mutual information is observed to be significantly influenced by the relative positions of the energy levels of $\mu_H$ and $\mu_L$ to $\varepsilon_Y$. For all of the three cases, $p(1, 0)$ approaches to unity for larger $U$, which is the dominant factor for the mutual information. On the other hand, $q(1, 0)$ is smaller for higher $\mu_H$ and $\mu_L$ relative to $\varepsilon_Y$, owing to the dominance by $q(0, 1)$ and $q(1, 1)$, which provides higher mutual information through Equation (8). The plateau in the region $U = 1~2$ observed in the curve for the condition $\mu_H = 3.1$, $\mu_L = 2.9$ can be explained as follows. Figure 6 plots $p(x, y)$ in dependence on $U$ for the condition $\mu_H = 3.1$, $\mu_L = 2.9$. As observed, the evolution of $p(0, 1)$, which is the dominant factor for the increase in the mutual information in the small-$U$ regime, by $U$ is damped around the region $U = 1~2$ because the electronic energy level $\varepsilon_Y + U$ approaches $\mu_H$ and $\mu_L$. This situation of the relation of energy levels starts the escape of the electron from the system dot out to its reservoirs, and allows the entrance of the electron into the detector dot from its higher reservoir level, in the existence of the electronic inter-dot Coulomb repulsion by $U$. By contrast, the conditions $\mu_H = 0.2$, $\mu_L = 0.4$ and $\mu_H = 1.1$, $\mu_L = 0.9$ exhibit no such an intermediate plateau in the evolution of mutual information by $U$, as observed in Figures 3–5, because $\varepsilon_Y + U$ is always higher than $\mu_H$ and $\mu_L$. After the point $U = 2$, where $\varepsilon_Y + U$ overlaps with $\mu_H$ and $\mu_L$, $p(1, 0)$ steeply increases and becomes the dominant factor to recover the increase in the mutual information until reaching unity. Incidentally, it is observed in Figure 6 that after the conversion of the gradients between $p(0, 1)$ and $p(1, 0)$ around $U = 1~2$, as explained above, $p(1, 0)$ overtakes $p(0, 1)$ exactly at the point $U = 4$, where the electronic potential energy difference $\mu_D - \varepsilon_X$ for the detector dot equals (the middle of $\mu_H$ and $\mu_L$) − $\varepsilon_Y$ for the system dot, and thus the absorbing driving forces into the dots are leveled.

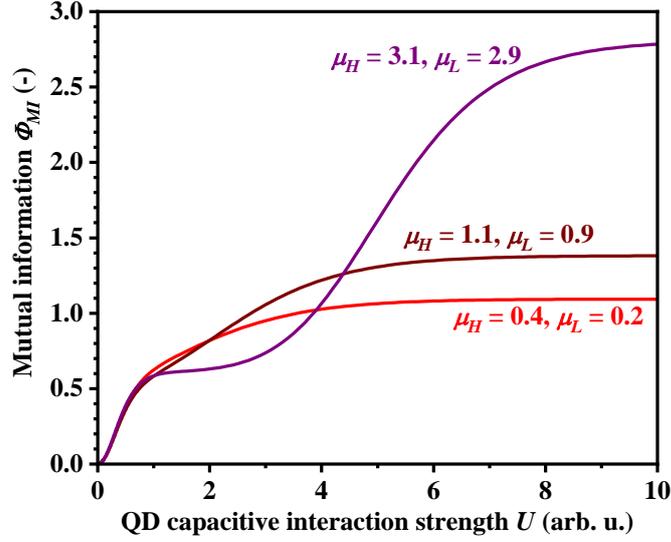

**Figure 5.** Mutual information $\Phi_{MI}$ in dependence on the capacitive interaction strength between the quantum dots $U$ for various relative energy-level positions $\mu_H$ and $\mu_L$ to $\varepsilon_Y$, keeping $\mu_H - \mu_L = 0.2$ constant, under the condition $\varepsilon_X = \varepsilon_Y = 1$, $T_D = 0.1$, $T_S = 1$, $\Gamma = 100$, $\Gamma_0^H = \Gamma_1^L = 10$, and $\Gamma_1^H = \Gamma_0^L = 0.1$.

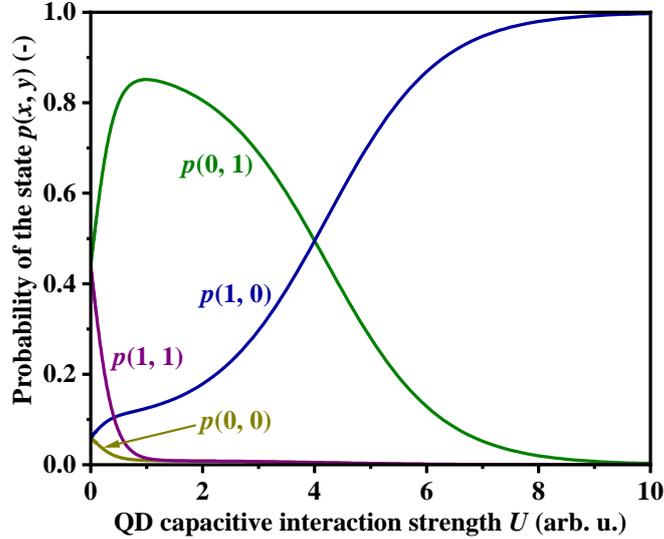

**Figure 6.** Probability of the state $p(x, y)$ in the coupled double quantum dot system in dependence on the capacitive interaction strength between the quantum dots $U$ under the condition $\varepsilon_X = \varepsilon_Y = 1$, $\mu_H = 3.1$, $\mu_L = 2.9$, $T_D = 0.1$, $T_S = 1$, $\Gamma = 100$, $\Gamma_0^H = \Gamma_1^L = 10$, and $\Gamma_1^H = \Gamma_0^L = 0.1$.

On a potentially common question if we should really consider the quantum dots conscious, we have to mention that it is difficult to answer at the present stage. One reason for the difficulty is that the integrated information theory has not been fully established or verified yet. Therefore, we have not yet understood if the integrated information, such as the Kullback–Leibler divergence between informationally connected and disconnected systems, indeed corresponds to the intensity of consciousness. Secondly, the answer to the question is thought to depend on the definition of the consciousness. For example, if the consciousness were inversely defined through the framework of the present integrated information theory or via the Kullback–Leibler divergence, yes the quantum dots can be considered conscious. We believe that at the point in the future where these two issues, the connection between the integrated information and consciousness and the definition of consciousness, have been clarified, we can answer the question. Until then, we would like to suspend any shallow conjecture. Although we employed a system comprising quantum dots, our numerical treatment for the master transition-rate equations in this study was classical. It may be interesting to explore the quantum-mechanical derivatives of our model setup, such as networks of qubits with their probability distributions replaced by density matrices, in our future study, as the attempts of quantum integrated information theory recently made [38–40].

## 4. Conclusions

In this study, we demonstrated a series of numerical simulations for the mutual information, as a measure of integrated information, in a model coupled double quantum dot system with an artificial function to "check" the electronic state of each other via the Coulomb interaction. The evolutions of the mutual information of the system in dependence on $U$ for various setup conditions were quantitatively analyzed in relation to the probability distributions of the electronic states. It was reasonably observed that the mutual information increases with $U$, because stronger interaction between the quantum dots provides a higher degree of integrated information. The mutual information of the system was observed strongly dependent on the relative temperatures of the detector and system sides owing to the influence of the electron-transport selectivity, while moderately insensitive to the antisymmetric relative amplitudes of the electronic tunneling rates. The mutual information is also found to be significantly influenced by the relative positions of the energy levels of $\mu_H$ and $\mu_L$ to $\varepsilon_Y$, due to the drastic turnover of the dominant electronic states. Our numerical model could be a simple but useful numerical tool for the future investigations of integrated information.

**Conflicts of Interest:** The author declares no conflict of interest.